\documentstyle[11pt,aaspp4,psfig]{article}
\begin{document}    
\newcommand{\be}{\begin{equation}}
\newcommand{\ba}{\begin{eqnarray}}
\newcommand{\ee}{\end{equation}}
\newcommand{\ea}{\end{eqnarray}}  

\title{Is There a Detectable Vishniac Effect?}
      
\author{Evan Scannapieco}

\affil{Departments of Physics and Astronomy,
University of California, Berkeley, CA 94720-7304} 

\begin{abstract}
The dominant linear contribution to cosmic microwave background (CMB)
fluctuations at small angular scales ($\lesssim 1'$) is a
second-order contribution known as the Vishniac or Ostriker-Vishniac
effect.  This effect is caused by the scattering of CMB photons off
free electrons after the universe has been reionized, and is dominated
by linear perturbations near the $R_V =2$ Mpc/($h \Gamma/0.2)$ scale
in the Cold Dark Matter cosmogony.  As the reionization of the
universe requires that nonlinear objects exist on some scale, however,
one can compare the scale responsible for reionization to $R_V$ and
ask if a linear treatment is even feasible in different scenarios of
reionization.  For an $\Omega_0 = 1$ cosmology normalized to cluster
abundances, only $\sim 65 \%$ of the linear integral is valid
if reionization is due to quasars in halos of mass
$\sim 10^9 M_\odot$, while $\sim 75\%$ of the integral is valid if
reionization was caused by stars in halos of $\sim 10^6 M_\odot$.  In
$\Lambda$ or open cosmologies, both the redshift of reionization and
$z_V$ are pushed further back, but still only $\sim 75 \%$ to $\sim 85
\%$ of the linear integral is valid, independent of the ionization
scenario.   We point out that all odd higher-order moments from
Vishniac fluctuations are zero while even moments are non-zero,
regardless of the gaussianity of the density perturbations.  This
provides a defining characteristic of the Vishniac effect that
differentiates it from other secondary perturbations and may be
helpful in separating them.

\end{abstract}

\newpage

\section{Introduction}

While recombination at $z \approx 1100$ marked the end of ionized
hydrogen from the viewpoint of a linearly evolving universe, the
nonlinear evolution of small-scale perturbations resulted in the
reionization of the intergalactic medium at much lower redshifts.  The
fact that quasar spectra show an absence of an absorption trough from
Ly$\alpha$ resonant scattering by neutral H atoms distributed
diffusely along the line of sight, the Gunn-Peterson effect (Gunn \&
Peterson 1965), means that this reionization must have occurred with a
high degree of efficiency before a redshift of 5.

One of the necessary consequences of this reionization is the presence
of secondary anisotropies in the cosmic microwave background (CMB)
due to the scattering of photons off ionized electrons.  These
secondary fluctuations can be divided into two classes: anisotropies
due to nonlinear structures and linear anisotropies.

Nonlinear secondary anisotropies are of several types.  Some of the
more studied of these include the scattering of photons off the hot
intracluster medium of galaxy clusters (Sunyaev \& Zel'dovich 1970,
1972; or for more recent treatments see, e.g., Evrard \& Henry 1991;
Colfrancesco et al.\ 1994; Aghanim et al.\ 1997), gravitational
lensing (see, e.g., Linder 1997; Metcalf \& Silk 1997), 
the impact of inhomogeneous reionization
(Aghanim et al.\ 1995; Peebles \& Juszkiewicz 1998; Knox, Scoccimarro,
\& Dodelson 1998), and the Rees-Sciama effect due to the bulk motions
of collapsing nonlinear structures (see, e.g., Rees \& Sciama 1968;
Kaiser 1982; Seljak 1996).

Small-scale linear anisotropies come in fewer flavors.  Detailed
analyses of linear perturbations have uncovered a single dominant
effect known as the Vishniac or Ostriker-Vishniac effect
(Hu, Scott, \& Silk 1994; Dodelson \& Jubas 1995; Hu \& White 1995; Hu
\& Sugiyama 1996).  The level of these perturbations has been
calculated by several authors (Ostriker \& Vishniac 1985; Vishniac
1987; Jaffe \& Kamionkowski 1998, hereafter JK).

These investigations raise the question of whether a detectable
Vishniac effect even exists since nonlinear structures must exist on
{\em some} length scale at the time of secondary scattering of CMB
photons, as it is only by the formation of nonlinear objects that the
universe is able to reionize itself.  If these scales are comparable
to those making the dominant contribution to the Vishniac effect, then
a linear analysis is inappropriate and a calculation of secondary
anisotropies must incorporate nonlinear effects.

In this work we determine the minimum length scale, $R_V$, which must
remain linear in order for a linear approach to scattering by ionized
regions with varying bulk motions to be accurate for the range of
angular scales over which one can hope to measure secondary fluctuations.
In hierarchical
scenarios of structure formation, such as the Cold Dark Matter (CDM)
model, smaller structures assemble at early times, later merging to
form larger objects.  This allows us to place limits on the time
between the formation of structures large enough to reionize the
universe and the time at which $R_V$ becomes nonlinear.
At that point, while peculiar
velocities of ionized gas continue to be imprinted on the microwave
background, the nature of this signature is qualitatively different
and is best interpreted from another perspective.

The structure of this work is as follows. In Sec.\ 2 we describe the
Vishniac effect and determine the physical length scale on which it
depends.  In Sec.\ 3 we compare this to the scale of reionizing
objects in different reionization scenarios and discuss the
applicability of linear theory.  In Sec.\ 4 we examine how
the Vishniac effect is distinguished from other effects.  Conclusions
are given in Sec.\ 5, and the various cosmological expressions used
throughout are summarized in the appendix.

\section{Analysis}

\subsection{Approximations}

The Vishniac effect is caused by the scattering of CMB photons
 by ionized regions with varying bulk motions.  The temperature 
fluctuations induced along a line of sight are given by 
\begin{equation}
\frac {\Delta T}{T}(\vec{\theta}) = 
- \int^{t_0}_0 dt \sigma_T e^{-\tau(\vec{\theta},t)} n_e(\vec{\theta},t) 
{\bf \hat{\theta}} \cdot {\bf v}(\vec{\theta},t), 
\end{equation}
where $\tau (\vec{\theta},t)$,
$n_e(\vec{\theta},t)$, and ${\bf v}(\vec{\theta},t)$  
are the optical depth along the line of sight, 
electron density, and bulk velocity,
$\sigma_T$ is the cross section for Thomson scattering,
$t$ is the age of the universe, and $t_0$ is the present age.
Following JK, we choose a coordinate system in which 
${\bf \hat{\theta}}$ represents a three-dimensional unit vector along the 
line of sight, $\vec{\theta}$ represents a two-dimensional unit
vector in the plane perpendicular to it, and bold letters represent fully 
three-dimensional vectors.  Thus ${\bf v} = (v_x,v_y,v_z)$,
$\vec{\theta} = (\theta_1, \theta_2, 0)$, and
${\bf \hat{\theta}} = (\theta_1, \theta_2, \sqrt{1 - \theta_1^2 - \theta_2^2})
\approx (\theta_1, \theta_2, 1)$, 
the validly of the approximation deriving from the small-scale
nature of the effect. Note that $n_e$, ${\bf v}$, and $\tau$ are all functions
of position, the optical depth being given by
$\tau(\vec{\theta},t) = \int^{t_0}_t
\sigma_T n_e(\vec{\theta},t) c dt' $, where $c$ is the speed of light.

If we decompose the density field into average and fluctuating
components, we obtain, to leading order,
\begin{equation}
\frac {\Delta T}{T}(\vec{\theta}) =
- \frac{\sigma_T n_0}{c} 
\int_0^1 dw a_0^3\frac{x_e({\bf \hat{\theta}} w_{\rm ang},w)}{a(w)^2}
(1+\delta({\bf \hat{\theta}} w_{\rm ang},w)-
\Delta \tau({\bf \hat{\theta}} w_{\rm ang},w))
{\bf \hat{\theta}} \cdot 
{\bf v}({\bf \hat{\theta}} w_{\rm ang},w) e^{-\tau_0(w)},  
\label{eq:deltat}
\end{equation}
where $x_e({\bf x},w)$ is the ionization fraction, $\tau_0$ and
$\Delta \tau({\bf x},w)$ are the optical depths due to the average and
fluctuating density components respectively, $a(w)$ is the scale
factor with $a_0 \equiv a(0)$, $\delta({\bf x},w)$ is the overdensity
field defined such that $\delta({\bf x},w) \equiv \rho({\bf
x},w)/\bar{\rho}(w) - 1$ where $\rho({\bf x},w)$ is the density field
and $\bar{\rho}(w)$ the average density as a function of comoving
distance, $n_0$ is the present average electron density, and $w_{\rm
ang}$ is the comoving angular distance, given by Eq.\
(\ref{eq:wangle}).  Note that we have replaced time by $w$, the
comoving distance defined by $dw \equiv c dt/a(t)$, and rewritten
${\bf v}$, and $\tau$ in comoving coordinates, ${\bf x}$.  Taking
the mass fraction of He to be $\sim 25 \%$, and approximating helium
reionization as simultaneous to that of hydrogen, $n_0 = \Omega_b
\rho_c /m_p \times 7/8 = 9.9 \times 10^{-6} \Omega_b h^2 {\rm
cm}^{-3}$, where $\rho_c$ is the critical density, $m_p$ is the mass
of the proton, $h$ is Hubble's constant normalized to $100 \, {\rm km}
\, {\rm s}^{-1} \, {\rm Mpc}^{-1}$, and $\Omega_b$ is the present
baryonic matter density in units of the critical density.

The Vishniac effect arises by considering the
homogeneously-reionizing, low optical depth case. In this case
$x_e({\bf x},w)$ is independent of position and $\Delta
\tau$ can be ignored.  In practice, realistic
reionization scenarios lead to low values of optical depth and hence
dropping $\Delta \tau$ is a safe assumption (Hu,
Scott, \& Silk 1994).  Indeed, the measurement of CMB fluctuations at
large scales precludes the high degree of damping that would be 
caused by a high optical depth 
(Scott \& White 1994; Hancock et al.\ 1998).  The
homogeneity of reionization, however, is a question of scales and thus
represents a basic assumption on which the Vishniac analysis is based.

In this limit Eq.\ (\ref{eq:deltat})
becomes
\be
\frac {\Delta T}{T}(\vec{\theta}) =
- \int_0^1 dw g(w) {\bf \hat{\theta}} \cdot 
\left( {\bf v}({\bf \hat{\theta}} w_{\rm ang},w) 
+ {\bf q}({\bf \hat{\theta}} w_{\rm ang},w) \right),
\label{eq:thin}
\ee
where 
${\bf q}({\bf x},t) \equiv {\bf v}({\bf x},w) \delta({\bf x},w)$
and $g(w)$ is the visibility function
\be
g(w) \equiv \frac{a_0^3 n_0 x_e(w)}{c a(w)^2} \sigma_T e^{-\tau}
= \frac{.121 \Omega_b h}{c} (1+z(w))^2 x_e(w) e^{-\tau},
\ee
with our conventions for the scale factor as in Appendix A.
This gives the probability of scattering off reionized electrons
and is a slowly-varying function of $w$.

Finally, we must approximate both ${\bf v}$ and ${\bf q}$ using linear
theory.  In this case,
the density contrast at a comoving coordinate ${\bf x}$ 
and comoving distance $w$ from the observer
is a random field with Fourier transform given by
$
  \tilde \delta({\bf k},w) \equiv \int \,d^3 {\bf x} \exp(-i
 {\bf k} \cdot {\bf x})\, \delta({\bf x},w).$
The spatial and time dependence of $\tilde \delta({\bf k},w)$
can be factorized, so $\tilde \delta({\bf k},w)=
\tilde \delta_0({\bf k})D(w)/D_0$
where $\tilde \delta_0({\bf k}) \equiv \tilde \delta({\bf k},0)$, 
$D(w)$ is the linear
growth factor, given by Eq.\ (\ref{eq:growth}), and $D_0 \equiv D(0)$.  
The power spectrum is then defined by the relation
\be
\langle \tilde{\delta_0}({\bf k}) \tilde{\delta_0}({\bf k'}) \rangle =
\langle \tilde{\delta_0}({\bf k}) \tilde{\delta_0}^*(-{\bf k'}) \rangle =
(2 \pi)^3 \delta^3({\bf k} + {\bf k'}) P(k),
\label{eq:deltadelta}
\ee 
where $\delta^3( {\bf k} + {\bf k'})$ denotes the
three-dimensional Dirac delta function.  This completely specifies the
probability density functional from which $\delta({\bf k})$ is drawn
in gaussian theories.  In the CDM cosmogony, $P(k)$
is given by Eq.\ (\ref{eq:pk}) and Eq.\ (\ref{eq:cdm}), and is
dependent on the `shape parameter' $\Gamma$, which is given as a
function of cosmological parameters by Eq.\ (\ref{eq:shapep}) and
constrained by observations of the galaxy correlation function to be
$0.23^{+0.042}_{-0.034}$ (Viana and Liddle 1996).  The overall
normalization of $P(k)$ can be fixed by the amplitude of mass
fluctuations on the 8 $h^{-1}$ Mpc scale as defined in Eq.\
(\ref{eq:sig}).

The linear velocity field is simply related to the density field by
the continuity equation ${\bf \nabla \cdot v}({\bf x}) 
= - a(w) {\dot {\delta}} ({\bf x},w)$ which in Fourier space gives
\be
     \tilde {\bf v}({\bf k},t) = {i a(w) \over k^2}\, {\dot D(w)
     \over D_0} \,{\bf k} \, \tilde \delta_0({\bf k}),
\label{eq:veckeqn}
\ee
where $\dot D(w)$ denotes the derivative of $D$ with respect to time
rather than $w$, and is given by Eq.\ (\ref{eq:growthdot}).
As the velocity always points along the direction of ${\bf k}$, only
${\bf k}$ modes with large values along the line of sight can have
large peculiar velocities in the ${\bf \hat{\theta}}$ direction.  But
these modes are varying with wavelengths much smaller than variations
in the window function and cancel when projected along the line of
sight.  It is only the second order (${\bf q}$) term then, that
contributes to the integral in Eq.\ (\ref{eq:thin}).

A simple real-space argument gives a different way of understanding
this cancellation.  As gravitational perturbations to a pressureless
fluid can be written in terms of the gradient of a scalar potential
field (${\bf v} \propto {\bf \nabla} \phi$ where $\phi$ is the
gravitational potential) the curl of the velocity field is zero to all
orders.  Thus a line integral of ${\bf v}$ along a line of sight is
approximately the integral of a gradient and is zero except for a
small contribution at the end points.  From all the terms
in Eq.\ \ref{eq:deltat}, only the 
$\delta \times  {\bf \hat{\theta}} \cdot {\bf v}$ survives to
contribute to $\frac{\Delta T}{T}$.

\subsection{Power Spectrum}

Dropping the velocity term from Eq.\ \ref{eq:thin},
we can analytically construct the
power spectrum of the angular fluctuations in the linear limit.
Let us define $\tilde{\frac{\Delta T}{T}}({\vec \kappa})$ as the
Fourier transform of the temperature fluctuations such that
$
     \tilde{\frac{\Delta T}{T}}({\vec \kappa}) 
= \int \,d^2 {\vec \theta} \exp(-i{\vec \kappa} \cdot {\vec \theta})\, 
	\frac{\Delta T}{T}({\vec \theta}),$
with the angular power spectrum defined as
$P_{\rm ang}(\kappa_1) (2 \pi)^2 
\delta^2({\vec \kappa_1} + {\vec \kappa_2}) \equiv \langle 
\left( \tilde{	\frac{\Delta T}{T}}({\vec \kappa_1}) 
       \tilde{  \frac{\Delta T}{T}}({\vec \kappa_2}) 
\right) \rangle$.
At the small angular scales appropriate to the Vishniac effect, 
$P_{\rm ang}(\kappa)$ is simply related to the usual $C_\ell$s 
used to express CMB fluctuations by $C_\ell = P_{\rm ang}(\kappa=\ell)$ 
(JK).

Several authors (Vishniac 1987; Kaiser 1992; JK) have derived
expressions for the Vishniac $C_\ell$s.  Here we provide a new
approach that, unlike other techniques,  is easily extended to 
calculate higher-order moments, as is shown in \S4.
Our method is a simple extension of
the usual formalism used to calculate single-point moments of the
$\frac{\Delta T}{T}(\vec \theta)$ distribution.

The simplest quantity of this sort is the second moment
$\langle \left[\frac{\Delta T_B}{T}(0)\right]^2\rangle$,  where we use
$B$ to denote convolution with a beam profile.
Given such a profile in Fourier space $B({\vec \kappa})$, 
the second moment can be calculated as
\be
\langle
\left[ \frac{\Delta T_B}{T}(0) \right]^2
\rangle = 
\int \frac{d^2 {\vec \kappa}}
{(2 \pi)^2}
 B({\vec \kappa})^2 
P_{\rm ang}(\kappa).
\label{eq:po}  
\ee
Suppose, however, that instead of asking about the second moment 
calculated from a single map observed with a beam 
$B({\vec \kappa})$, we instead compute the single-point function as 
calculated from the convolution of two maps, observed by
beams $B_1({\vec \kappa})$ and $B_2({\vec \kappa})$.  As these profiles
are arbitrary, we are free to take
\be
B_1({\vec \kappa}) \equiv (2 \pi)^2
\delta^2({\vec \kappa}-{\vec \kappa_1}) 
\qquad \qquad
B_2({\vec \kappa}) \equiv (2 \pi)^2
\delta^2({\vec \kappa}-{\vec \kappa_2}),
\label{eq:deltawin}
\ee
where $\delta^2({\vec \kappa})$ is the two-dimensional delta function.
In this case we find
\be
\langle
\frac{\Delta T_{B_1}}{T}(0)
\frac{\Delta T_{B_2}}{T}(0)
\rangle = (2 \pi)^2 P_{\rm ang}(\kappa_1) 
\delta^2({\vec \kappa_1}+ {\vec \kappa_2}),
\ee
recovering the angular power spectrum.
Thus the power spectrum of the Vishniac effect can be computed
from the single-point correlation if we are careful to
express the beam profiles in sufficient generality.

Let us consider then
\ba
\langle
\frac{\Delta T_{B_1}}{T}(0)
\frac{\Delta T_{B_2}}{T}(0) 
\rangle = 
& &
\int_0^1 dw_1 g(w_1) \int_0^1 dw_2 g(w_2)
\int \frac{d^3{\bf k_a}}{(2 \pi)^3} \int \frac{d^3{\bf k_b}}{(2 \pi)^3}
\nonumber \\ & &
B_1({\vec k_a} w_{1,{\rm ang}})
B_2({\vec k_b} w_{2,{\rm ang}})
e^{i k_{a,z} w_1 + i k_{b,z} w_2}
\langle
\tilde{q}_z({\bf k_a},w_1) \tilde{q}_z({\bf k_b},w_2) 
\rangle,
\label{eq:b1b2}
\ea
where $\tilde{\bf q}({\bf k},w)$ is the Fourier transform of
${\bf q}({\bf x},w)$.
Substituting in the expression for $\tilde{\bf q}$ in terms of
$\tilde{\delta}$:
\be
{\bf {\tilde q}}({\bf k},w) = 
\frac{ i a(w) \dot D(w) D(w)}{D_0^2} 
\int \frac{d^3 {\bf k'}}{(2 \pi)^3} 
\tilde{\delta}_0({\bf k'})
\tilde{\delta}_0({\bf k} - {\bf k'}) \frac{{\bf k'}}{k'^2},
\ee
Eq.\ (\ref{eq:b1b2}) becomes
\ba
\langle
\frac{\Delta T_{B_1}}{T}(0)
\frac{\Delta T_{B_2}}{T}(0) 
\rangle = 
&-& 
\int_0^1 dw_1 G(w_1)
\int_0^1 dw_2 G(w_2)
\prod_{j=1}^4 
\left[
\int \frac{d^3  {\bf k_j}}{(2 \pi)^3}  
\right] B_1(({\vec k_1}+{\vec k_2}) w_{1,{\rm ang}})
\nonumber \\
& &
B_2(({\vec k_3}+{\vec k_4}) w_{2,{\rm ang}}) 
e^{i (k_1+k_2) w_1+ i (k_3+k_4) w_2}
\frac{k_{1,z}}{k_1^2}
\frac{k_{3,z}}{k_3^2} 
\langle
\prod_{l=1}^4 \tilde{\delta}_0({\bf k_l})
\rangle,
\ea
where
\be
G(w) \equiv \frac{ g(w) a(w) D(w) \dot D(w)}{D_0^2}.
\label{eq:bigg(w)}
\ee
As $G(w)$ is slowly varying, we can follow Kaiser (1992) in dividing
the integrals over comoving distance into $N$ statistically
independent intervals of width $\Delta w$, over each of which $G(w)$ is
well approximated by a constant.  In this case
\ba
\langle
\frac{\Delta T_{B_1}}{T}(0)
\frac{\Delta T_{B_2}}{T}(0) 
\rangle = 
-\sum_{n=1}^{N}
G(w_n)^2 \Delta w^2
\prod_{j=1}^4 
\left[
\int \frac{d^3  {\bf k_j}}{(2 \pi)^3}  
\right] \qquad \qquad 
\nonumber \\  
 B_1(({\vec k_1}+{\vec k_2})w_{n,{\rm ang}})
 B_2(({\vec k_3}+{\vec k_4})w_{n,{\rm ang}})
\nonumber \\  
j_0 \left(\frac{(k_{1,z}+k_{2,z}) \Delta w}{2} \right)
j_0 \left(\frac{(k_{3,z}+k_{4,z}) \Delta w}{2} \right)
\frac{k_{1,z}}{k_1^2}
\frac{k_{3,z}}{k_3^2}
\langle
\prod_{l=1}^4 \tilde{\delta}_0({\bf k_l})
\rangle.
\ea
As the density fluctuations are taken to be gaussian, we can expand the 
expectation value of the product of overdensities by Wick's theorem, 
keeping only the terms in which $k_1$ is paired with $k_3$ or $k_4$.
If we then define $k'_2  \equiv k_1 + k_2$, we find
\ba
\langle
\frac{\Delta T_{B_1}}{T}(0)
\frac{\Delta T_{B_2}}{T}(0) 
\rangle = 
\sum_{n=1}^{N}
G(w_n)^2 \Delta w^2
\int \frac{d^3{\bf k_1}}{(2 \pi)^3}
\int \frac{d^3{\bf k_2'}}{(2 \pi)^3}
B_1({\vec k'_2} w_{n,{\rm ang}})
B_2(-{\vec k'_2} w_{n,{\rm ang}}) \nonumber \\
\left[ \frac{k_{1,z}^2}{k_1^4} + \frac{k_{1,z} 
(k'_{2,z} - k_{1,z})}{k_1^2 || {\bf k_2'} -  {\bf k_1} ||^2}
\right]
P(k_1) P(|| {\bf k'_2} - {\bf k_1} ||) 
j_0^2 \left(\frac{ k'_{2,z} \Delta w}{2} \right).
\ea
The Bessel function has a width $\delta k_z \sim 1/\Delta w$
so $k_{2,z} << \kappa/w_{\rm ang}$ wherever $j^2_0$ is appreciable.
We can neglect terms that are smaller by a factor of 
$w_{\rm ang}^2 \kappa^2/\Delta w^2$ to obtain 
\ba
\langle
\frac{\Delta T_{B_1}}{T}(0)
\frac{\Delta T_{B_2}}{T}(0) 
\rangle  & = &
\sum_{n=1}^{N} G(w_n)^2 \Delta w 
\int \frac{d^3{\bf k_1}}{(2 \pi)^3}
\int \frac{d^2{\vec k'_2}}{(2 \pi)^2} 
B_1({\vec k'_2}  w_{\rm ang})
B_2(-{\vec k'_2} w_{\rm ang}) \nonumber \\
& &
\frac{k_{1,z}^2}{k_1^4} - 
\frac{k_{1,z}^2}{k_1^2 || ({\vec k_2'},0) -  {\bf k_1} ||^2}
P(k_1) P(|| ({\vec k'_2},0) - {\bf k_1} ||).
\ea
Taking 
\be
B_1({\vec \kappa}) =
B_2({\vec \kappa}) = 
2 \pi \sigma^2 e^{-\frac{\sigma^2 \kappa^2}{2}}
\ee
gives the second moment as observed by a beam of gaussian width
$\sigma$, while choosing beam profiles as given in Eq.\
(\ref{eq:deltawin}) yields the angular power spectrum.  In this case
\be
C_\ell = P_{\rm ang}(\kappa = \ell) 
= \int_0^1 dw \frac{G(w)^2}{w^2_{\rm ang}} 
P_{\rm V}(\ell / w_{\rm ang}),
\label{eq:clcalc}
\ee
where
\be
P_{\rm V}({k}) =
\int \frac{d^3{\bf k_1}}{(2 \pi)^3}
P(k') P(|| ((k,0,0) - {\bf k_1} ||) 
\left[
\frac{k_{1,z}^{'2}}{k_1^{'4}} 
- \frac{k_{1,z}^{'2}}{k_1^{'2} ||(k,0,0) - {\bf k_1} ||^2 } 
\right],
\ee
which, choosing a coordinate system in which the $z'$ axis points
along the direction of $(k,0,0)$ becomes
\be
P_{\rm V}(k) =
\frac{k}{8 \pi^2} \int_0^\infty dx \int_{-1}^{1} 
d \mu P(k x) P(k \sqrt{1 - 2 x \mu + x^2})
(1 - \mu^2) 
\left[1 - 
\frac{x^2}{1 - 2 x \mu + x^2}
\right].
\ee
This is equivalent to the usual expression for the Vishniac power spectrum
\be
P_{{\rm V}} (k) = 
\frac{k}{8 \pi^2} 
\int_0^\infty dx \int_{-1}^{1} d \mu P(k x) 
P( k \sqrt{ 1  - 2 x \mu + x^2}) 
\frac{ (1 - \mu^2)(1 - 2 x \mu)^2}{(1 - 2 x \mu + x^2)^2},
\label{eq:pperp}
\ee
as can be seen by rewriting both integrals in rectangular coordinates
and applying an origin shift.
Thus the $C_\ell$s  are dependent on an integral along the line of
sight of a term, $P_{\rm V}(k)$, that is independent of redshift and
arises from the convolution of the ${\bf q}$ fields.
We note in passing that our results are in agreement with 
Dodelson and Jubas (1995) and are twice the values found in JK.

\subsection{Physical Scales}

\begin{figure}
\centerline{\psfig{file=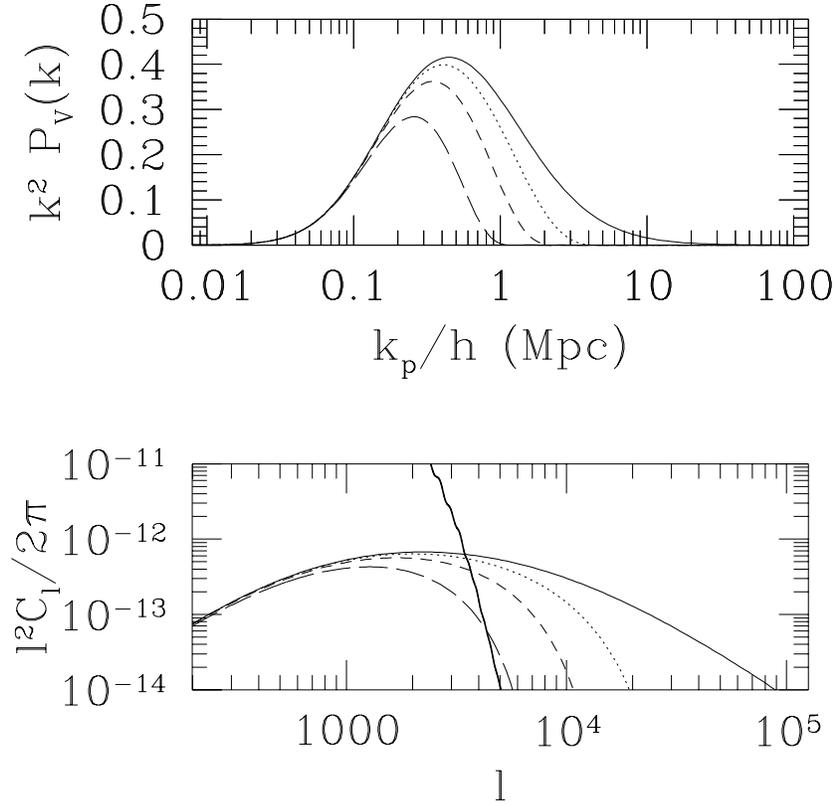,width=4.5in}}
\caption{
Upper panel: $P_{\rm V} (k)$ from $\Lambda$CDM density
fluctuations convolved with top-hat window functions of various
scales.  The solid line is the exact analytic expression, the dotted
line corresponds to a top-hat filtering scale of $R$ = 1 Mpc/h, the
short-dashed to $R$ = 2 Mpc/h, and the long dashed to $R$ = 4 Mpc/h.
The overall normalization is arbitrary as it is dependent on the
choice of the scale factor.  Lower panel: $\ell^2 C_\ell/2\pi$ from density
fluctuations convolved with top-hat window functions of various
scales.  Window functions are as in the upper panel.
The sharply falling solid line shows the primary anisotropies 
as calculated by CMBFAST.}
\label{fig:sk}
\end{figure}

Having outlined the approximations which are used to calculate
this effect and constructed the resulting power spectrum of
fluctuations,
we now examine which physical scales contribute most to $P_{\rm V}(k)$.
Typically, Eq.\ (\ref{eq:clcalc})
is used to calculate the Vishniac effect by integrating over a
particular matter power spectrum.  To study the dependence of the
effect on physical scale, we replace the power spectrum that appears
in Eq.\ (\ref{eq:pperp}) with $P(k) W^2(k R)$ where $P(k)$ is the CDM
power spectrum and $W(x)$ is the spherical top-hat window function,
given by Eq.\ (\ref{eq:tophat}).  Here we consider a $\Lambda$CDM
model in which the current nonrelativistic matter, vacuum, and
baryonic densities in units of the critical density are $\Omega_0$ =
0.35, $\Omega_\Lambda$ = 0.65, and $\Omega_b$ = 0.06 respectively, and
the ``tilt'' in the power spectrum as parameterized in Eq.\
(\ref{eq:pk}) is taken to be flat, $n = 1.0$.  If the other parameters
are taken to be $h$ =0.65, $\sigma_8$ = 1.05, $\Gamma = 0.2$ and
reionization occurs instantaneously and completely at $z_{\rm re} =
18$, this results in $P_{\rm V}(k)$ and $C_\ell$ as plotted in Fig.\
\ref{fig:sk}.  Note that these values correspond to $\frac{\Delta T}{T}$s 
a full order of magnitude smaller than large-scale primary
anisotropies, pointing out the experimental challenges that must be
overcome before secondary anisotropies can be measured (see, e.g.,
Subrahmanyan et al.\ 1993; Church et al.\ 1997).  For reference
we also plot the COBE-normalized primary fluctuations as computed by 
the CMBFAST code V2.4.1
(Seljak \& Zaldarriaga 1996; Hu et al.\ 1998; Zaldarriaga \& Seljak 1998),
for the same cosmological model.
Note that the position of this line is sensitive only to the Silk
damping scale and thus is largely independent of the shape of the
primordial power spectrum.

In this figure we see that even when filtered at the $1$ Mpc/$h$
scale, the power spectrum is appreciably changed, especially at higher
wavenumbers.  While low $\ell$ fluctuations are not affected,
a comparison with the solid line shows that these fluctuations
are lost in the CMB primary signal and are very difficult to measure.
Note that the damping shown in this graph represents not
only a loss of power in linear theory, but also an increase of
nonlinear power.  Thus, at the point that the $1$ Mpc/$h$ scale has
become nonlinear, the Vishniac effect is competing with nonlinear
effects over the range of angular scales in which it is able to be detected.

By the time the $2$ Mpc/$h$ scale becomes nonlinear, however, the peak
wavelength of the Vishniac effect has been shifted by a factor of
$1/2$ and $\sim 50 \%$ of the power of $P_{\rm V}(k)$ has been lost.
At this point, linear calculations are unlikely to be 
reliable at measurable $\ell$ values $\gtrsim 4000$, and a more
careful theoretical approach becomes necessary.
These length scales are inversely proportional to the `shape
parameter' $\Gamma$ which is $0.2$ for this model.  Thus we can
conservatively fix $R_V =$ 2 Mpc/$h \Gamma_{0.2}$ where
$\Gamma_{0.2} \equiv \Gamma/0.2$, as the maximum nonlinear length
scale that still allows a linear analysis to be appropriate.

\section{Redshift of applicability}

Having determined $R_V$, we now consider what scenarios of
reionization are compatible with a Vishniac effect.  These scenarios
can be roughly divided into two classes: those in which the dominant
source of ionizing photons is due to stars formed in dwarf galaxies
with halo masses $\gtrsim 10^6 M_\odot$ (Couchman \& Rees 1986;
Fukugita \& Kawakasi 1994; Shapiro, Giroux, \& Babul 1994; Haiman \&
Loeb 1997) and models in which reionization occurs due to active
nuclei in galaxies with halo masses $\gtrsim 10^9 M_\odot$ (Efstathiou
\& Rees 1988; Haehnelt \& Rees 1993; Aghanim et al.\ 1995; Haiman \&
Loeb 1998; Valageas \& Silk 1999).  See also, however, the issues
raised in Madau, Haardt, \& Rees (1998) and Miralda-Escud\'e,
Haehnelt, \& Rees (1998), and the more exotic scenarios of
reionization described in Scott, Rees, \& Sciama (1991) and Adams,
Sarkar, \& Sciama (1998).

Models in which smaller objects are the most important predict
redshifts of reionization of $z_{\rm re} \approx 20$ while models in
which reionization is due to objects on scales $\gtrsim 10^9 M_\odot$
predict more modest values of $z_{\rm re}$.  All models of
reionization, however, are constrained by the lack of a Gunn-Peterson
absorption trough in the spectra of high-redshift quasars, implying
that the intergalactic medium was highly deficient in neutral hydrogen
at redshifts $\lesssim 5.$ Thus, high-mass reionization scenarios can only be
successful in cosmologies in which the parameters are such that
relatively large objects became nonlinear at high redshifts, precisely
those models where the linear approximation is on the most shaky
ground.

Using our value for $R_V$ from Sec.\ 2, we
are able to determine a redshift of applicability, $z_V$, below
which the Vishniac approximation is invalid over the measurable
range of $\ell$ scales.  
The number density of objects above a critical over-density,
$\delta_c$, is given by Press-Schechter theory 
(Press \& Schechter 1974) as
\be
\frac{d n(M,z)}{dM} = -\sqrt{\frac{2}{\pi}} \frac{\rho(z)}{M}
\frac{\delta_c D_0}{\sigma(M)^2 D(z)} 
\exp \left(
-\frac{\delta_c^2 D_0^2}{2 \sigma^2(M) D(z)^2}
\right) \frac{d \sigma}{dM}(M),
\ee
where $n(M,z)$ is the number density of collapsed objects per unit mass at a
redshift of $z$, $\rho(z)$ is the comoving density of the universe at
a redshift of $z$, $D(z)$ is the linear growth factor of fluctuations,
and $\sigma(M)$ is the level of fluctuations on the mass scale
corresponding to a sphere containing a mass $M$, which can be computed
from Eq.\ (\ref{eq:sig}).  Typically, this formula is used to
determine the number density of virialized halos.  In this case
$\delta_c(z)$ is a weak function of $z$ for open models and $\Lambda$
models and a fixed value of $3 (12 \pi)^{2/3}/20 \simeq 1.69$ in the
$\Omega_0 = 1$ case (Kitayama \& Suto 1996).

As a rough rule of thumb we can assume that reionization 
takes place when the 2$\sigma$ fluctuations at the relevant 
scale have collapsed.  In this case,
$D(z_{\rm re})/D_0 = \delta_c(z_{\rm re})/ (2 \sqrt{2} \sigma(M))$ which is
$\approx 0.6 /\sigma(M)$ in the flat case.  We take the linear approximation
to be valid up the point at which the 1$\sigma$ scale fluctuations at the
$R_V$ scale have reached an overdensity of 1.  This gives 
$D(z_V)/D_0 = 1/(\sqrt{2} \sigma(R_V)) \approx 
0.7/\sigma(R_V).$  

\begin{figure}
\centerline{\psfig{file=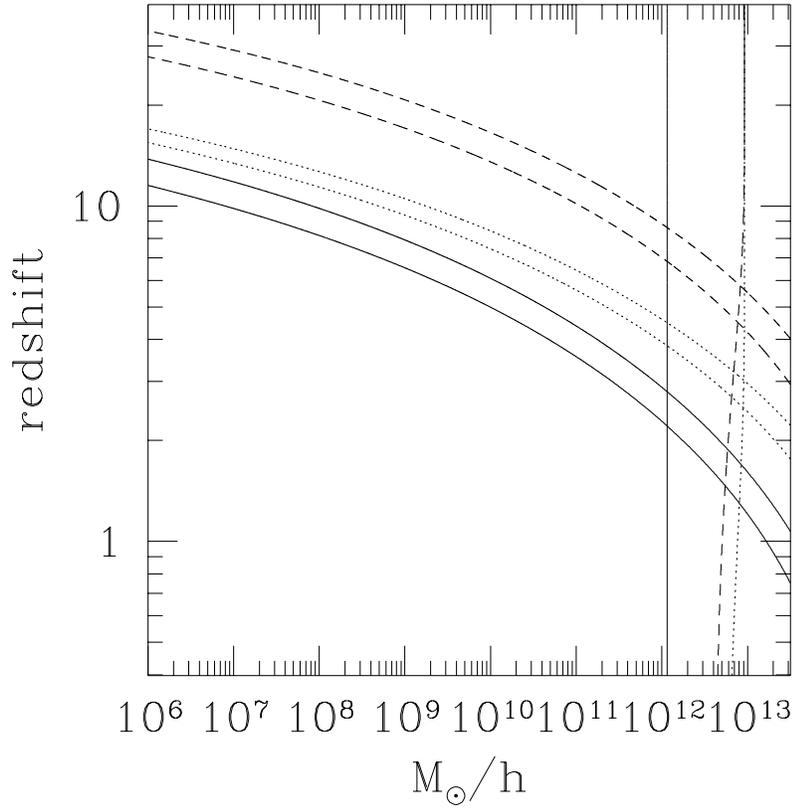,width=4.5in}}
\caption{
In each pair of horizontal lines the upper 
correspond to the ionization redshift
$D(z_{\rm re})/D_0 = \delta(z_{\rm re})/ (2 \sqrt{2} \sigma(M))$, and 
the lower lines correspond to the redshift at which a linear treatment
is no longer valid, $D(z_V)/D_0 = 1/(\sqrt{2}\sigma(M))$.
The solid lines correspond to the flat model, the dotted lines to
the ``concordance model,'' and the dashed lines to the open model,
with parameters as described in the text.  The nearly vertical lines show the
mass scales corresponding to spheres of radius $R_V$ for each of the
models. }
\label{fig:sigmas}
\end{figure}

In Fig.\ \ref{fig:sigmas} we plot both $z_{\rm re}$ and $z_V$ as
functions of mass, as it is the mass scale rather than the length scale
that is most easily identified with different reionization scenarios.
We consider three cosmologies, representative of
parameters that favor both low mass-scale and high mass-scale
scenarios of reionization.  In order to compare with a scenario that
is representative of stellar reionization, we consider a flat model
normalized at the 8 Mpc/$h$ scale (Viana \& Liddle 1996).  Here
$\Omega_0$ = 1.0, $\Omega_\Lambda$ = 0.0, $\Omega_b$ = 0.07, $h$ =
0.5, $\sigma_8$ = 0.60, and $n$ = 1.0.  In this
scenario $\Gamma = 0.44$, shifting the CDM line and decreasing $R_V$
by a factor of $\sim 2$.  Note that this value of $\Gamma$ is
incompatible with the observed galaxy correlation function.  Typical
of high-mass reionization scenarios, we consider the ``concordance
model'' of Ostriker \& Steinhart (1995), which was used by Haiman \&
Loeb (1998a) in their modeling of reionization by quasars.  In this
case the parameters are taken to be $\Omega_0$ = 0.35,
$\Omega_\Lambda$ = 0.65, $\Omega_b$ = 0.04, $h$ = 0.65, $\sigma_8$ =
0.87, $\Gamma = 0.20$, and $n$ = 0.96.  Finally, we examine an open
model with $\Omega_0 = 0.35$, again normalized at 8 Mpc/$h$.  In this
case $\Omega_\Lambda$ = 0.0, $\Omega_b$ = 0.04, $h$ = 0.65, $\sigma_8$
= 1.02, $\Gamma = 0.20$, and $n$ = 1.0.

Let us first consider the $\Omega_0 = 1$ model, represented by the lowest pair
of lines.  In this model $\sigma_8$ is the lowest of all 
the cosmologies considered and the perturbations evolve the most quickly.
The combination of these two effects moves the 
redshift of applicability down to a value of $z_V \approx 2$,
and thus one might expect to find an appreciable Vishniac effect.
The problem, however, is that the low
normalization and rapid evolution of perturbations also lowers the
collapse redshift of the objects responsible for reionization. In this 
scenario, the $2 \sigma$, $10^9 M_\odot$ peaks collapse at
$z_{\rm re} \approx 7$.  As reionization must have occurred with a high degree 
of efficiency by $z = 5$, this scenario is only marginally 
consistent with quasar absorption-line observations.  Thus a flat cosmology 
is most compatible with low-mass reionization scenarios.
If the collapse of $10^6 M_\odot$ halos is responsible for reionization, then
$z_{\rm re} \approx 13$ for this model, yielding a larger range
of redshifts over which the Vishniac effect could be imprinted on the microwave
background.  This redshift is comparable to the revised values calculated
in the stellar reionization scenario of Haiman and Loeb (1997; 1998b), although
they consider a somewhat higher range of $\sigma_8$ values.

\begin{figure}
\centerline{\psfig{file=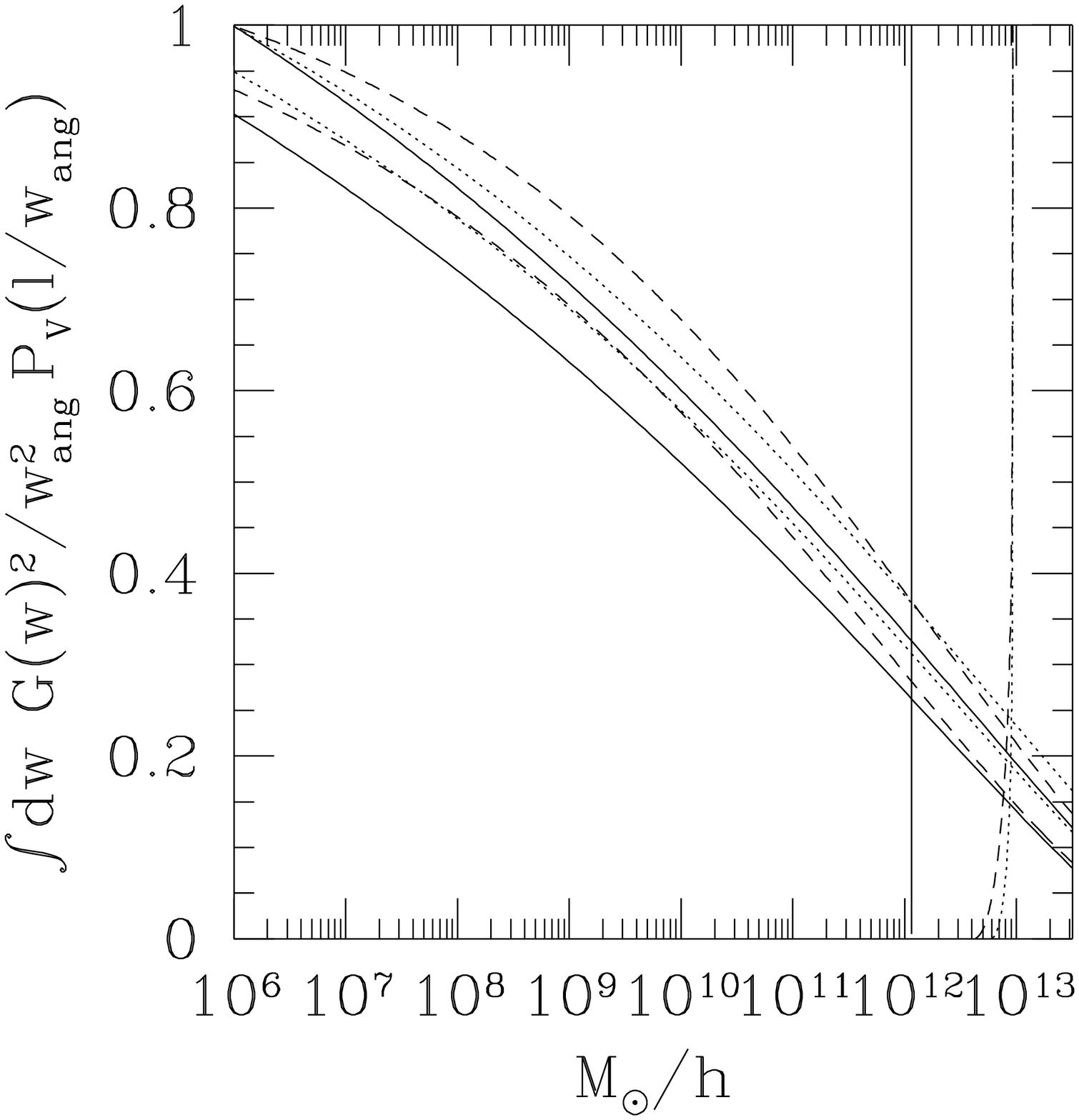,width=3.2in}}
\caption{
Fig.\ \protect\ref{fig:sigmas} replotted in terms of
$\int_0^{w(z(M))} dw G(w)^2/w_{{\rm ang}}^2 P_V(\ell/w_{{\rm ang}})$ 
with $\ell = 4000$ for three different cosmologies. Lines are as in 
Fig.\ \protect\ref{fig:sigmas}.}
\label{fig:width1}
\end{figure}

\begin{figure}
\centerline{\psfig{file=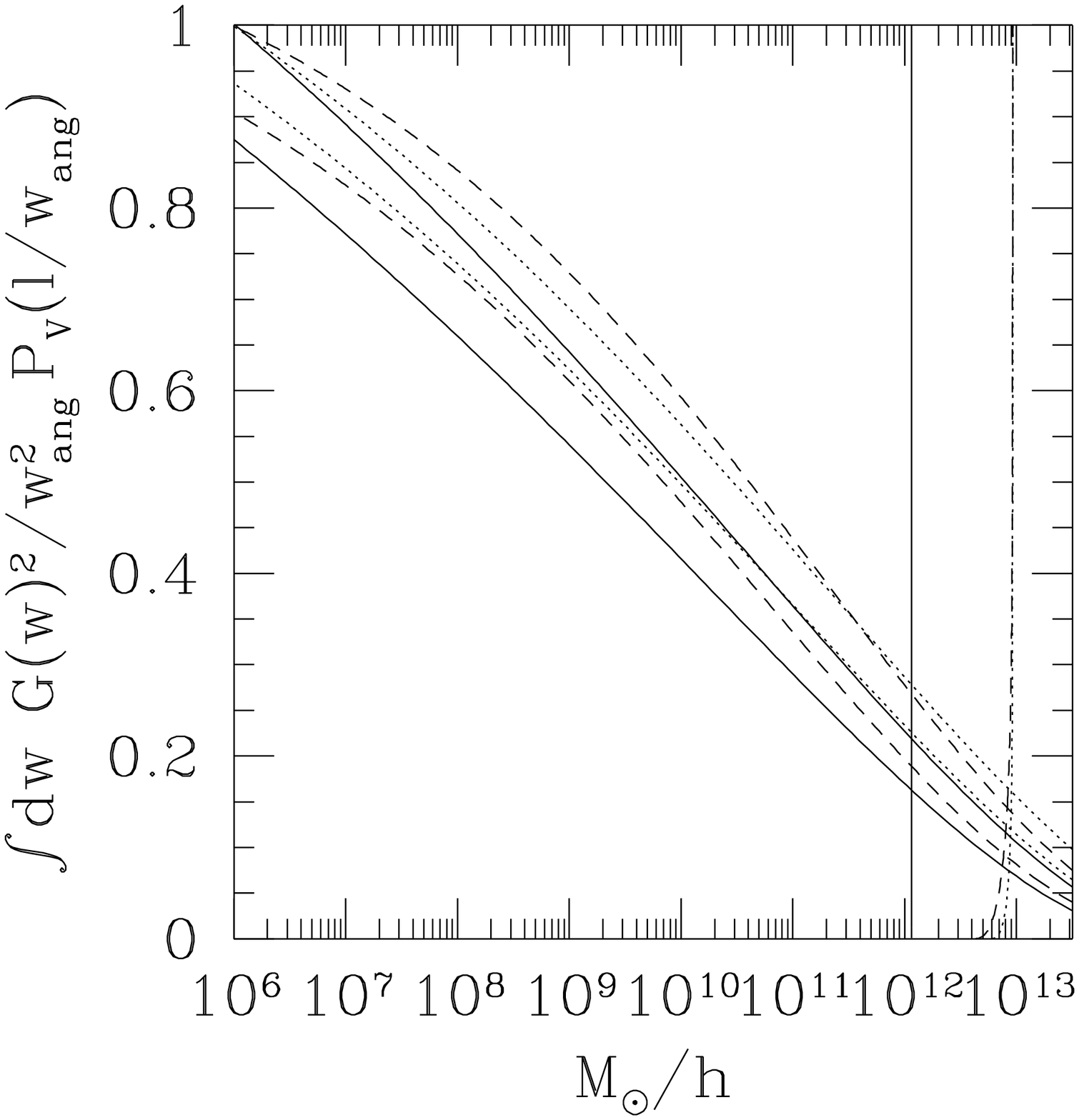,width=3.2in}}
\caption{
Same as Fig.\ \protect\ref{fig:width1} but with $\ell = 12000$.  
The onset of nonlinearity at the $R = 1 \, {\rm Mpc}(h \Gamma/0.2)$ 
scale can be estimated by shifting the nearly vertical lines to
masses lower by a factor of 8.}
\label{fig:width2}
\end{figure}

To quantify this further, 
in Figs.\ \ref{fig:width1} and \ref{fig:width2} we replot 
Fig.\ \ref{fig:sigmas}, replacing the vertical axis with   
$\int_0^{w(z)} dw G(w)^2/w_{{\rm ang}}^2 P_V(\ell/w_{{\rm ang}})$,
the contribution to $C_\ell$ due to bulk motions within
a redshift of $z$.  We take $\ell = 4000$ in Fig.\ \ref{fig:width1}
and $\ell = 12000$ in Fig.\ \ref{fig:width2}, and normalize
the $y$ axis such that the integral is equal to 1 at a redshift of 
$z_{\rm re}(10^6 M_\odot)$.  The magnitude of the Vishniac $C_\ell$ is 
then directly proportional to the vertical width of the gap between 
$\int_0^{w(z_{\rm res})} dw 
G(w)^2/w_{{\rm ang}}^2 P_V(\ell/w_{{\rm ang}})$, and
$\int_0^{w(z_V)} dw 
G(w)^2/w_{{\rm ang}}^2 P_V(\ell/w_{{\rm ang}})$,
allowing us to judge the linear and nonlinear 
contributions to Eq.\ (\ref{eq:clcalc}) in arbitrary scenarios of 
reionization at a glance.

From this point of view, reionization by $10^6 M_\odot$ objects
results in only a marginal improvement the accuracy of a linear
treatment.  While 35\% of the $\ell = 4000$ integral is nonlinear if
$M_{\rm re} = 10^9 M_\odot$, the $M_{\rm re} = 10^6 M_\odot$ case is
still 25\% inaccurate.  These numbers are somewhat lower in the
high-$\ell$ case, in which 18\% of the integral is nonlinear in the
low mass case, and 25\% in the high mass.  Note however, that our
definition of $R_V = 2 \, {\rm Mpc}(h \Gamma/0.2)$ was based on the
damping of perturbations at $\ell = 4000.$ From Fig.\ \ref{fig:sk} we
see that perturbations at $\ell = 12000$ are largely damped when the
matter power spectrum is filtered at the $R = 1 \, {\rm
Mpc}(h \Gamma/0.2)$ scale, and a more fair comparison between Figs.\
\ref{fig:width1} and \ref{fig:width2} would be to shift the vertical
lines in Fig.\ \ref{fig:width2} to masses lower by a factor of 8,
yielding much the same numbers as in the $\ell = 4000$ case.

More typical of high-mass reionization scenarios is the $\Lambda$CDM
model represented by the dotted lines in Figs.\
\ref{fig:sigmas}-\ref{fig:width2}.  In this model, $\sigma_8$ is
slightly higher than in the flat case and the evolution of $D(z)$ is
slowed.  These effects raise the collapse redshift of $10^9 M_\odot$
peaks to $z_{\rm re} \approx 11$, easily compatible with Gunn-Peterson
tests.  Note that our crude estimate of the redshift of reionization
is almost the same as the redshift of $\approx 12$ calculated
by Haiman \& Loeb (1998a) for the same set of cosmological parameters,
using a more sophisticated Press-Schechter based argument for the
ionizing flux from quasars.  In this scenario $z_V$ is also pushed
back, although this is lessened by the shift of $\Gamma$ as compared
to the flat case.  Thus $z_V \approx 2.5$.

One might imagine that in this case, a much wider margin of 
redshifts would lead to an accurate linear calculation.
Fig.\ \ref{fig:width1} indicates otherwise.  
In this case almost 20\% of the low-mass and over 25\% of the 
high-mass integral comes from redshifts at which $R_V$ is nonlinear.  
The high-$\ell$ values are slightly lower, with 12\% of 
the $10^6 M_\odot$ and 17\%  of the $10^9 M_\odot$ integral taking 
place when $R_V$ is nonlinear, but again these numbers become
roughly the same as the $\ell = 4000$ case for a more fair comparison.

The most extreme case we consider is the cluster-normalized open
model, in which $\sigma_8$ is the highest and $D(z)$ the most slowly
evolving.  In this case $z_V \approx 4$, the $10^9 M_\odot$ and the
$10^6 M_\odot$ schemes reionize at $z_{\rm re} \approx 22$ and $z_{\rm
re} \approx 33$ respectively.  This cosmology yields the largest
regime of redshift space over which a linear analysis is valid and the
most accurate results. Here nonlinear $R_V$ scale fluctuations
contaminate 15\% of the low-mass and 20\% of the high-mass $C_{4000}$
integrals.  Again these numbers are lower at higher
$\ell$ but roughly the same after accounting for the smaller filtering
scale of $R = 1 \, {\rm Mpc}(h \Gamma/0.2)$.

\begin{figure}
\centerline{\psfig{file=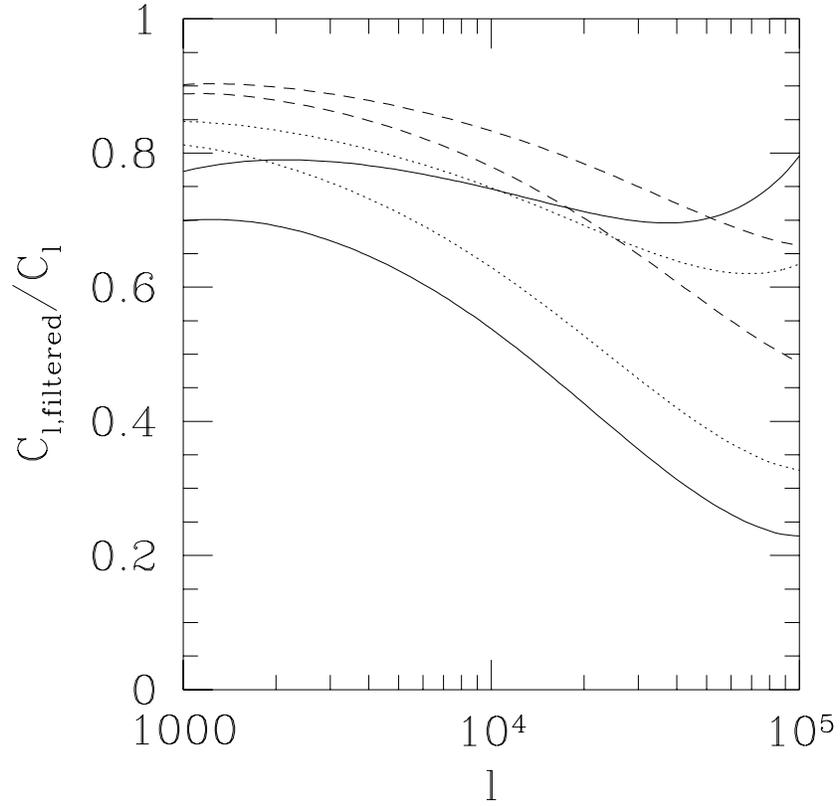,width=4.5in}}
\caption{
Ratio of the contribution to $C_\ell$ from the linear regime to the total
calculated contribution as a function of $\ell$.
The solid lines correspond to the flat model, the dotted lines to
the ``concordance model,'' and the dashed lines to the open model,
and in each pair of lines the upper line corresponds to reionization
by $10^6 M_\odot$ objects and the lower line to reionization by
$10^9 M_\odot$ objects.}
\label{fig:nailincofin}
\end{figure}

As a final check of the validity of our analysis, we construct
$C_{\ell,{\rm filtered}}$ defined as the angular power spectrum as given
by Eq.\ (\ref{eq:clcalc}) but replacing $P(k)$ with $P(k) W^2(k R_{\rm nl}(z))$
where $R_{\rm nl}(z)$ is now the nonlinear length scale {\em at each redshift}
as in Fig.\ 2 ($D(z)/D_0 =0.7/\sigma(R_{\rm nl}(z))$).  In Fig.\
\ref{fig:nailincofin} we plot the ratio of $C_{\ell,{\rm filtered}}$ 
to $C_\ell$ calculated from the unfiltered CDM power spectrum.  While
only a few reionization scenarios are represented in this graph, this
nevertheless gives us some feel of the accuracy of the linear treatment
over different scales, and unlike Figs.\ 2 - 4, is completely
independent of our definition of $R_V$.  Here we see that
in the range of $\ell$
values at which the effect is most likely to be measured 
($4000 \lesssim \ell \lesssim 120000$) this estimate is in good agreement
with the accuracies given in the previous figures.  Note also
that a linear treatment becomes increasingly inaccurate with $\ell$,
and thus measurements of fluctuations at angular scales just below
this Silk damping scale will be most easy to interpret.

From these results we can safely conclude that even given our present
ignorance as to the cosmological parameters, no more than $\approx 85
\%$ of the contribution to Eq.\ (\ref{eq:clcalc}) for measurable
$\ell$ values can be calculated by linear theory in a CDM cosmogony.  This
depends only on the shape of the CDM power spectrum and the lack of 
diffuse L$\alpha$ absorption in quasar spectra out to $z \gtrsim 5$.

\section{Is There a Detectable Vishniac Effect, Really?}

At this point, one may raise the objection that our argument is a bit
semantic, as there will still be scattering due to bulk motions even
when linear theory breaks down.  Indeed, the
Kinetic Sunyaev-Zel'dovich effect, which is due to the peculiar
velocities of clusters, can be viewed as a nonlinear counterpart
to the Vishniac effect.  Is there not, then, a sort of
detectable Vishniac effect
in cosmological scenarios in which $R_V$ is nonlinear during the
epoch of reionization, albeit under a different name?

The problem with this point of view is that it overlooks some
important physical distinctions between these effects.  The Vishniac
effect is due to the presence of a redshift regime in which $G(w)$ is
slowly varying and a delicate cancellation takes place due to the lack
of curl in the peculiar velocity field.  As it can be calculated
precisely, it provides a unique probe of the reionization history of
the universe that is not available from measurements of nonlinear
fluctuations.

Furthermore, the Vishniac effect displays a distinct signature
of higher-order moments that allows it to be distinguished from 
other contributions. In order to understand why this occurs,
let us consider the bispectrum $B({\vec \kappa_1},{\vec \kappa_2}) 
(2 \pi)^2 \delta^2({\vec \kappa_1} + {\vec \kappa_2} +  
	{\vec \kappa_3}) \equiv
\langle
\tilde{\frac{\Delta T}{T}}({\vec \kappa_1})
\tilde{\frac{\Delta T}{T}}({\vec \kappa_2}) 
\tilde{\frac{\Delta T}{T}}({\vec \kappa_3})
\rangle$.
We can apply our $\delta$-function beam approach in
the linear regime to calculate this as
\ba
\langle
\tilde{\frac{\Delta T}{T}}({\vec \kappa_1})
\tilde{\frac{\Delta T}{T}}({\vec \kappa_2}) 
\tilde{\frac{\Delta T}{T}}({\vec \kappa_3})
\rangle 
 = -i
\prod_{i=1}^3 
\left[\int_0^{w_0} dw_i G(w_i) \right]
\prod_{j=1}^6 
\left[
\int \frac{d^3  {\bf k_j}}{(2 \pi)^3}  
\right]
\nonumber \\  \qquad (2 \pi)^6
 \delta^2(({\vec k_1}+{\vec k_2})w_{1,{\rm ang}}-{\vec \kappa_1})
 \delta^2(({\vec k_3}+{\vec k_4})w_{2,{\rm ang}}-{\vec \kappa_2})
 \delta^2(({\vec k_5}+{\vec k_6})w_{3,{\rm ang}}-{\vec \kappa_3})
\nonumber \\  \qquad \qquad
e^{i(k_{1,z}+k_{2,z})w_1 + i (k_{3,z}+k_{4,z})w_2 + i(k_{5,z}+k_{6,z})w_3}
\frac{k_{1,z}}{k_1^2}
\frac{k_{3,z}}{k_3^2}
\frac{k_{5,z}}{k_5^2}
\langle
\prod_{l=1}^6 \tilde{\delta_0} ({\bf k_l},w_i)
\rangle.
\label{eq:bispec}
\ea
If $G(w)$ is slowly varying,
we can again follow Kaiser (1992) in dividing
the integrals over comoving distance into $N$ statistically
independent intervals of width $\Delta w$, 
\ba
\langle
\tilde{\frac{\Delta T}{T}}({\vec \kappa_1})
\tilde{\frac{\Delta T}{T}}({\vec \kappa_2}) 
\tilde{\frac{\Delta T}{T}}({\vec \kappa_3})
\rangle
 = -i
\sum_{n=1}^{N}
G(w_n)^3 \Delta w^3
\prod_{j=1}^6 
\left[
\int \frac{d^3  {\bf k_j}}{(2 \pi)^3}  
\right] (2 \pi)^6
 \delta^2(({\vec k_1}+{\vec k_2})w_{n,{\rm ang}}-{\vec \kappa_1})
\nonumber \\  
 \delta^2(({\vec k_3}+{\vec k_4})w_{n,{\rm ang}}-{\vec \kappa_2})
 \delta^2(({\vec k_5}+{\vec k_6})w_{n,{\rm ang}}-{\vec \kappa_3})
j_0 \left(\frac{(k_{1,z}+k_{2,z}) \Delta w}{2} \right)
\nonumber \\  
j_0 \left(\frac{(k_{3,z}+k_{4,z}) \Delta w}{2} \right)
j_0 \left(\frac{(k_{5,z}+k_{6,z}) \Delta w}{2} \right)
\frac{k_{1,z}}{k_1^2}
\frac{k_{3,z}}{k_3^2}
\frac{k_{5,z}}{k_5^2}
\langle
\prod_{l=1}^6 \tilde{\delta_0} ({\bf k_l},w_n)
\rangle.
\label{eq:bispec2}
\ea
As there are an odd number of $k_z$ terms, there is no pairing of density
fields that does not result in an odd $k_z$ term.  As all the $k$ 
integrals are even, 
$\langle \tilde{\frac{\Delta T}{T}}({\vec \kappa_1}) 
 	 \tilde{\frac{\Delta T}{T}}({\vec \kappa_2}) 
	 \tilde{\frac{\Delta T}{T}}({\vec \kappa_3}) \rangle $ = 0.

This cancellation can be understood from a more general perspective.
$B({\vec \kappa_1},{\vec \kappa_2})$ is generated
by the expectation value of the triple product of the field ${\bf q}$.
As ${\bf q}$ is an isotropic vector field, 
$\langle \tilde{q}_{i}({\bf k_1}) \tilde{q}_{j}({\bf k_2}) 
	 \tilde{q}_{k}({\bf k_3}) \rangle$
can depend on no vectors other than the ${\bf k}$ vectors themselves
and must therefore be proportional to at least one of 
the them (Monin \& Yaglom 1971).  This means that 
the $i = j = l = z$ component of this product must be proportional
to $k_{1,z}$, $k_{2,z}$, or $k_{3,z}$.  Thus the 
odd $k_z$ term that results in $B({\vec \kappa_1},{\vec \kappa_2}) = 0$ 
is due to the isotropy of the
density fluctuations.  By a similar argument, all odd moments 
of the temperature fluctuations must be zero as these also depend on the
expectation value of the product of an odd number
of ${\bf q}$'s. Note that this is true independent of the gaussianity
of the probability distribution functional of ${\tilde \delta}({\bf k})$.

This cancellation does not apply to the even moments, however, as an
even number of ${\bf q}$'s can be arranged in a way that is not proportional
to one of the ${\bf k}$ vectors.  Thus the Vishniac effect is unique in that
it is nongaussian independent of the gaussianity of the probability
distribution functional of $\tilde{\delta}({\bf k})$, but this
nongaussianity is expressed only in the even higher-order moments.
This alternation of zero-and-nonzero higher-order moments provides a
unique signal that distinguishes the Vishniac effect from other
secondary anisotropies, and provides us with the opportunity to use
nongaussian statistics as a discriminator between these contributions.
Note however, that the difficulty of measuring secondary anisotropies
may make such an analysis difficult to apply in practice.

\section{Conclusions}

Due to the tremendous predictive power of linear theory, comparisons
between linear predictions and large-scale cosmic microwave background
measurements promise to constrain cosmological parameters to the order
of a few percent (Jungman et al.\ 1996; Bond, Efstathiou, \& Tegmark
1997; Zaldarriaga, Spergel, \& Seljak 1997).  The natural extension of
this approach is to try to measure small-scale secondary anisotropies
and match them to linear predictions to study the reionization history
of universe.  The situation in this case is more muddled, however, as
a number of nonlinear secondary effects also contribute at these
scales.

The dominant secondary linear anisotropy is a
second-order contribution known as the Vishniac or Ostriker-Vishniac
effect.  As this effect can be predicted accurately as a function of
cosmological parameters, several authors have proposed that its
measurement will prove to be a sensitive probe of the reionization
history of the universe.  Reionization occurs by the formation of
nonlinear structures, however, raising the question of whether a
regime of redshift space exists in which these objects have collapsed
but a linear analysis is still appropriate.

In this work, we have determined the relevant physical scales that
give rise to the Vishniac effect in a Cold Dark Matter cosmogony,
showing that approximations are already compromised when $1$ Mpc/($h
\Gamma_{0.20}$) scales have become nonlinear, and break down when $2$
Mpc/($h \Gamma_{0.20}$) dark matter halos reach overdensities of 1.
The width of the redshift regime over which the effect can be
imprinted on the CMB is dependent on the cosmological parameters and
the reionizing mass scale.  Schemes in which reionization is due to
radiation from active galactic nuclei associated with dark matter
halos of masses $\gtrsim 10^9 M_\odot$ are limited by the absence of a
Gunn-Peterson absorption trough.  As reionization must have occurred
with a high degree of efficiency before a redshift of 5, such models
are successful only if one assumes a large value of $\sigma_8$, or
considers open models with slowly-changing linear growth factors.
Both these assumptions push back the redshift at which $R_V$ becomes
nonlinear, limiting the range over which a linear analysis is
appropriate.

Scenarios in which reionization is due to much smaller objects, such
as stars formed in dwarf galaxies associated with dark matter halos of
masses $\gtrsim 10^6 M_\odot$, are able to reionize the universe at
much larger redshifts even in cosmologies in which $\sigma_8$ is small
and $D(z)$ quickly evolving.  This represents only a marginal gain
however, as the high redshift contribution to the 
Vishniac integral is roughly proportional to comoving
distance, and comoving distances are 
small at high redshifts.  Thus low-mass scenarios of reionization are
more compatible with a linear analysis not so much because they
reionize earlier as because they allow $R_V$ to become nonlinear more
recently without violating Gunn-Peterson limits.

The Vishniac effect arises from physical processes that are distinct
from nonlinear secondary anisotropies.  Its detection indicates the
presence of a redshift regime in which a delicate cancellation takes
place due to the lack of curl in the peculiar velocity field and slow
variations in $G(w)$.  This leaves a unique signature in the
higher-order moments of the temperature fluctuations that is absent
from its nonlinear counterparts.  Furthermore, due to the predictive
power of linear theory, it represents a sensitive probe of the
reionization history not available from measurements of nonlinear
contributions.

As with measurements of large angular scale anisotropies, small-scale
microwave background anisotropy measurements have the potential to
uncover much about the history of our universe.  Also as with
large-scale measurements, whether this potential will be realized
remains to be seen.  While the Vishniac effect represents a possible
probe of the reionization epoch, the analysis will, as always, be more
involved than first suggested.  Ultimately it will only be through the
measurement and analysis of small-scale microwave background
anisotropies that we will be able to know if there is a detectable Vishniac
effect.

\acknowledgments I wish to thank Nabila Aghanim, Fran\c cois Bouchet,
Rychard Bouwens, Andrew Jaffe, Douglas Scott, and Naoshi Sugiyama for
helpful discussions and am particularly indebted to Joseph Silk, whose
comments and suggestions have been invaluable during the preparation
of this work.  I thank Uro\v{s} Seljak and Matias Zaldarriaga for the use of
CMBFAST and acknowledge partial support by the NSF.

\section*{Appendix}

In this appendix, we provide explicit expressions for the
cosmological factors used throughout this paper.  We allow both 
$\Omega_0$ and $\Omega_\Lambda$ to be free. 
In this case the Friedman equations for the evolution of 
the scale factor of the Universe, $a(z)$ are
\begin{equation}
     {\dot a \over a} = H_0 E(z) \equiv H_0 \sqrt{\Omega_0 (1+z)^3 +
     \Omega_\Lambda + (1-\Omega_0-\Omega_\Lambda)(1+z)^2},
\end{equation}
and
\begin{equation}
     {\ddot a \over a} = H_0^2 [\Omega_\Lambda - \Omega_0
     (1+z)^3/2],
\end{equation}
where $H_0 = 100\, h$ km~sec$^{-1}$~Mpc$^{-1}$ is the Hubble
constant, and the overdot denotes a derivative with respect to
time.  

We choose the scale factor such that $a_0 H_0=2 c$.  If we are
located at the origin, $w = 0$, then an object at redshift $z$
is at a comoving distance,
$w(z) = {1\over 2} \int_0^z \, dz' E^{-1}(z')$
and at a time
$t(z) = {1 \over H_0} \int_z^\infty \, dz' (1+z')^{-1} E^{-1}(z').$
Note that this is different than conformal time,
defined by $d\eta=dt/a$, such that  
$c \eta(z) = w(\infty)-w(z)$.  
For any flat universe the angular size distance $w_{\rm ang} = w$,
and in an open universe
\be
 w_{\rm ang}  = \frac{\sinh(2 w \sqrt{1 - \Omega_0 - \Omega_\Lambda})}
                 {2   \sqrt{1 - \Omega_0 - \Omega_\Lambda}} .
\label{eq:wangle}
\ee
The growth factor as a function of redshift is 
\begin{equation}
     D(z) = {5 \Omega_0\, E(z) \over 2} \, \int_z^\infty \, {1+z' \over
     [E(z')]^3} \, dz',
\label{eq:growth}
\end{equation}
while  
\be
     {\dot D \over D} = {\ddot a \over \dot a} - {\dot a
     \over a} + {5 \Omega_0 \over 2} {\dot a \over a} {(1+z)^2
     \over [E(z)]^2\,D(z)}.
\label{eq:growthdot}
\ee
The evolution of $\Omega$ is given by
\be
\Omega(z) = \Omega_0 (1+z)^3 E^{-2}(z)
\ee
where $\Omega_0 = \Omega(0).$

For the power spectrum, we use
\begin{eqnarray}
      P(k) &=& {2 \pi^2 \over 8} \delta_H^2 (k/2)^n T^2(k_p\,{\rm
     Mpc}/h \Gamma),
\label{eq:pk}
\end{eqnarray}
where $T(q)$ is the CDM transfer function, $k_p= k/a_0 =k
H_0/2 c$ is the physical wavenumber, and
the shape parameter $\Gamma$ is defined as (Sugiyama 1995)
\be
\Gamma \equiv \Omega_0 (h/0.5) \exp(-\Omega_b-\Omega_b/\Omega_0).
\label{eq:shapep}
\ee
The factor of 8 in the denominator in
Eq.\ (\ref{eq:pk}) arises because we are using $a_0
H_0=2 c$.  For the transfer function, we use the analytic fit given 
by Bardeen et al.\ (1986) for the CDM cosmogony,
\begin{equation}
     T(q)= {\ln(1+2.34q)/(2.34 q)\over [1+3.89 q + (16.1 q)^2 +
     (5.46 q)^3 + (6.71 q)^4]^{1/4}}.
\label{eq:cdm}
\end{equation}
The normalization factor $\delta_H$ can be fixed by specifying 
$\sigma(8{\rm Mpc}/h)$, where
the variance of the mass enclosed in a sphere of radius $R$ is given by    
\be
\sigma^2(R) = \frac{1}{2 \pi^2} \int_0^\infty k^2 dk P(k) W^2(k_p R).
\label{eq:sig}
\ee
Here $W(x)$ is the spherical top-hat window function, defined in
Fourier space as
\be
W(x) \equiv 3 \left[ \frac{\sin(x)}{x^3} - \frac{\cos(x)}{x^2} \right].
\label{eq:tophat}
\ee

\fontsize{9}{11pt}\selectfont

\newpage

\end{document}